\documentclass[aps,prd,reprint,nofootinbib,longbibliography,showkeys]{revtex4-1}

\pdfoutput=1

\usepackage{blindtext}
\usepackage{mathtools}
\usepackage{xcolor}

\usepackage{dsfont}
\usepackage{siunitx}

\usepackage{mathrsfs}
\usepackage{braket}

\usepackage[utf8]{inputenc}
\usepackage[english]{babel}
\usepackage{hyperref}

\hypersetup{
    colorlinks=true,
    linkcolor=blue,
    filecolor=blue,      
    urlcolor=red,
    citecolor=blue,
}
\usepackage{color,colortbl}

\usepackage{tensind}
\tensordelimiter{?}

\usepackage{graphicx}
\usepackage{hyperref}
\DeclareGraphicsExtensions{.bmp,.png,.jpg,.pdf}
\usepackage{amsmath}
\usepackage{courier}
\usepackage{verbatim}
\usepackage{amssymb}
\usepackage{amsfonts}
\usepackage{natbib}

\newcommand{\dif}{\mathrm{d}}
\newcommand{\imathnew}{\mathrm{i}}
\newcommand{\be}{\begin{equation}}
\newcommand{\ee}{\end{equation}}
\begin{document}
\title{Radiation-dominated bouncing model with slow contraction and inflation }

\author{Piero A. P. Molinari$^{1}$, Paola C. M. Delgado$^{2}$, Rodrigo F. Pinheiro$^{1}$, Nelson Pinto-Neto$^{1}$}
\email{pieromolinari@cbpf.br}
\email{paola.moreira.delgado@doctoral.uj.edu.pl}
\email{rfpinheiro@cbpf.br}
\email{nelsonpn@cbpf.br}

\affiliation{$^1$ Centro Brasileiro de Pesquisas F\'{\i}sicas - CBPF, Rua Dr. Xavier Sigaud 150, Urca, CEP-22290-180, Rio de Janeiro, RJ, Brasil\\
$^2$Faculty of Physics, Astronomy and Applied Computer Science, Jagiellonian University, 30-348 Krakow, Poland
}

\begin{abstract}

A very simple non-singular inflationary model is presented where the unique matter content is a radiation fluid. The model slowly contracts from a very large, almost empty and flat spacetime and realizes a bounce. It is then launched to a quasi-de Sitter inflationary expansion with more than sixty e-folds, which smoothly changes to the usual classical, decelerated radiation-dominated expansion before nucleosynthesis. The initial contracting and final expanding phases are classical, but the intermediate bounce and inflationary phases are induced by quantum cosmological effects emerging from a Gaussian wave function quickly moving in configuration space. During this quantum era, a huge number of photons is created.
The scale factor describing all this rich evolution is a surprisingly simple analytic function of conformal time. The cosmological scalar perturbations arising from quantum vacuum fluctuations in the far past of the model present an almost scale invariant spectrum with an amplitude compatible with observations for reasonable values of the free parameters of the model. 
\end{abstract}

\maketitle

\section{Introduction}

Inflation is a simple idea which not only solves some important puzzles arising in Big Bang Cosmology \cite{guth}, but which has also predicted from weak assumptions the red tilted, almost scale invariant spectrum of scalar cosmological perturbations \cite{mukhanov}, afterwards confirmed by detailed observations of the cosmic microwave background (CMB) \cite{Planck:2018jri}. However, inflationary models generally make use of a scalar field called inflaton, which is not observed in Nature (except in the case of Higgs inflation \cite{higgs}); they also do not address the initial singularity of the Standard Cosmological Model (SCM) \cite{vilenkin}, turning the resulting scenario imcomplete in this sense. 

In order to solve this problem, bouncing models without singularities emerged in the last decades as possibilities to complete the SCM \cite{novello,melnikov,brandenberger,Pinto-Neto:2013toa,Pinto-Neto:2018zvn,Pinto-Neto:2021gcl}. In fact, such models do not contain the puzzles inherent to the Big Bang Cosmology \cite{PhysRevD.78.063506,Pinto-Neto:2021gcl}, and some of them can also lead to almost scale invariant spectra of scalar cosmological perturbations, although only in some specific cases. Some bouncing models have a unique bounce, usually with a matter dominated contraction in order to yield the correct spectrum of perturbations \cite{wands,peter,Cai:2011zx}, and others are cyclic, with a very slow contracting phase \cite{cyclic}. Usually, bouncing models do not require an inflationary phase \cite{PhysRevD.78.063506}, but they are not incompatible with inflation, with many scenarios containing both phases \cite{lqc}. In many cases, the bounce itself helps in yielding initial conditions for inflation. However, contrary to inflation, where the physical requirements on the inflaton field Lagrangian are standard, the bounce itself requires some new physics. It can come either from non-minimal couplings \cite{novello}, semi-classical corrections leading to nonlinear curvature terms in the gravitational action and/or an effective energy-momentum tensor of matter violating the null energy condition \cite{melnikov}, or quantum corrections arising from quantum gravity approaches applied to Cosmology \cite{Pinto-Neto:2013toa,lqc}.

In this paper, we present a cosmological model where the bounce and an expanding inflationary phase are both induced by quantum effects, without any inflaton field. The quantum effects arise from a canonical quantization of gravity restricted to the mini-superspace configuration of homogeneous and isotropic geometries, with a constant perfect fluid equation of state parameter $w=p/\rho$ satisfying $-1/3<w<1$, leading to an effective Schr\"odinger equation for the cosmological wave function. As is well known, this simple quantum geometrical approach can be viewed as an approximation of a more involved quantum theory of gravity, hence we limit the maximum energy density and spacetime curvature of the effective model to be some few orders of magnitude below the Planck scale, where this simple approach can be reliable \cite{kiefer}. The wave function solution is interpreted using the de Broglie-Bohm quantum theory \cite{Bohm:1951xw,PhysRev.85.180}, where a quantum scale factor evolution can be calculated. 

The initial wave function is chosen to be a Gaussian moving in configuration space. The time-dependent solution is calculated, leading to a bouncing non-singular asymmetric scale factor, reaching its standard classical evolution in the asymptotic past and future of the model. Depending on the sign of the momentum of the wave function, the classical contraction can be either much slower or faster than the classical expansion, with a huge creation or annihilation of particles between these phases, respectively.
In the sequel, it is noticed that, specifically for $w\approx 1/3$, the model experiences either a quantum inflationary or deflationary era during its evolution. Clearly, the physically interesting possibility is the former, on which we focused our attention.

The resulting model is a radiation-dominated universe classically and slowly contracting from an almost flat spacetime up to a quantum bounce, followed by an era of a quantum quasi-de Sitter expanding phase, which changes smoothly to the classical radiation-dominated era before nucleosynthesis. In order to have inflation during a sufficiently long period, the wave function from which it originates must be moving with high momentum in configuration space. During the quantum era, a large amount of radiation is quantum created.
Therefore, surprisingly, we obtain a very simple radiation-dominated, non-singular cosmological model, the primordial era of which combines the three main ingredients of the primordial Universe that have been investigated so far: a slow contraction \cite{cyclic,Cook:2020oaj} (although in a very different way, with possible different consequences), a bounce and inflation. The scalar power spectrum of cosmological perturbations is calculated, presenting an almost scale invariant behavior and an amplitude compatible with observations for very reasonable parameter choices; namely, that the minimum curvature scale of the model ($1/R^{1/2}$, where $R$ is the Ricci scalar) is around four orders of magnitude bigger than the Planck length. \\

The paper is divided as follows: in Section \ref{secBackEv} we obtain the wave function which originates the class of scale factors we will investigate. The free parameters are connected to physically meaningful cosmological quantities. A very simple expression for the scale factor and the independent free parameters of the theory is exhibited, and its properties in different phases are described in detail. It is shown that the case $w\approx 1/3$ implies an era of quantum quasi-de Sitter expansion after the bounce. In Section \ref{secanalytical} we make analytic estimations of the primordial scalar perturbations originated from quantum vacuum fluctuations in the asymptotic past of the model, first for general $w$, then for $w\approx 1/3$, calculating their amplitude and spectral index. In Section \ref{secnumerical}, the numerical calculations are shown, confirming the analytical estimations: an almost scale invariant spectrum of scalar perturbations, with an amplitude compatible with observations if the minimum curvature scale of the model is around three orders of magnitude larger than the Planck length. We conclude in Section \ref{conc} with a discussion and the possible future developments of the model.

\section{Background evolution}
\label{secBackEv}

We consider a homogeneous and isotropic universe filled with a perfect fluid governed by a barotropic equation of state of the form $p=w\rho$, with constant $w$. The Einstein-Hilbert action may be written using the ADM variables \cite{PhysRev.116.1322}, considering a foliation of spacetime into space-like hypersurfaces. It can be shown that, neglecting boundary terms, the corresponding FLRW action reads
\begin{align}
 S=&\frac{1}{6}\int\dif^4 x\sqrt{h}N\left[K_{ij}K^{ij}-\left(h^{ij}K_{ij}\right)^2+{}^{(3)}R\right]\nonumber\\
 =&\int N \dif t\left(-\frac{a\dot{a}^2}{N^2}+a\mathcal{K}\right)\,.
 \label{FLRW_action}
\end{align}
The quantities $N$, $h_{ij}$, $K_{ij}$,  ${}^{(3)}R$, $a$, and $\mathcal{K}$ are the lapse function, induced metric of the spacelike hypersurfaces, extrinsic curvature of the spacetime foliation, the Ricci scalar of the spacelike hypersurfaces, the scale factor, and the constant curvature of the homogeneous and isotropic spacelike hypersurfaces, respectively. An overdot corresponds to a derivative with respect to coordinate time $t$. The gravitational Hamiltonian follows directly from \eqref{FLRW_action} (we consider hereafter spatially flat hypersurfaces). 

On the other hand, the matter contribution can be obtained by means of the Schutz formalism \cite{PhysRevD.2.2762}, which then leads to the following total Hamiltonian:
\be
    \mathcal{H}=N \mathcal{H}_0=N\left(-\frac{P_a^2}{4a}+\frac{P_{T}}{a^{3w}}\right)\,.
\ee

The dynamical variables are the scale factor $a$, its conjugate momentum $P_a$, the perfect fluid variable $T$, related to the velocity field of the fluid, and its conjugate momentum $P_T$, related to the classically conserved total number of particles of the fluid. The latter appears linearly in the Hamiltonian, which leads to a natural interpretation of $T$ as a time variable. The cosmic time $\tau$ is related to $T$ through Hamilton's equations by $\dif \tau \equiv N\dif t=a^{3w}\dif T$. Note that for dust ($w=0$) $T$ is cosmic time, while in the case of a radiation fluid ($w=1/3$) $T$ corresponds to conformal time $\eta$. 
In order to perform the canonical quantization of the system, one must specify an operator ordering, which in this case is such that we obtain a covariant Laplacian under redefinitions of the scale factor $a$ \cite{Pinto-Neto:2021gcl,Pinto-Neto:2018zvn}. With this choice, the Wheeler-DeWitt equation $\hat{\mathcal{H}}_0\Psi=0$, where $\Psi$ denotes the minisuperspace wave function, reads
\begin{equation}\label{wdw equation}
\imathnew \frac{\partial }{\partial T}\Psi(\chi,T)=\frac{1}{4} \frac{\partial^{2}}{\partial \chi^{2}}\Psi(\chi,T)\,,
\end{equation}
where 
\begin{equation}
\chi \equiv \frac{2}{3(1-w)} a^{3(1-w)/2}\,.
\label{chi and a}
\end{equation}
The concrete solutions for the scale factor are obtained by proposing an \textit{ansatz} for the initial wave function, which must be propagated to lead to the wave function at any time $T$. \\

Let us turn our attention to the conceptual implications of this quantization procedure for a system that is the universe as a whole. The standard Copenhagen interpretation demands an external classical domain in order to collapse the wave function, as a result of a measurement process performed on the system. This external domain is by definition absent in a cosmological setting. One of the alternative interpretations of quantum mechanics that allows for consistent cosmological scenarios is the de Broglie-Bohm (dBB) Quantum Theory \cite{Bohm:1951xw, PhysRev.85.180}, which we shall adopt from now on. This approach consists of a deterministic interpretation of quantum mechanics in which the particles or field amplitudes describe trajectories in configuration space which are objectively real, regardless of a measurement process. These trajectories satisfy judiciously chosen guidance equations, in which the initial particle positions or field configurations are not known - only a probability distribution thereof -, thus constituting the hidden variables of the theory. If this probability distribution is given by the Born rule, then all probabilistic predictions of quantum theory are recovered. The compatibility of this interpretation with quantum cosmology is a result of the so called effective collapse, which describes the occupation of one of the branches of the wave function in a measurement process by the point particle in configuration space, depending on its initial particle positions or field configurations. Since this effective collapse does not require an external ``observer'', the dBB theory can be applied to the universe as whole. Moreover, the ``particle trajectory'', which in this case is related to the evolution of the scale factor, is part of an objective reality. The resulting cosmological models might be able to avoid the initial singularity problem, replacing it with a bounce, which is preceded by a contracting phase and followed by the usual expansion of the universe \cite{Pinto-Neto:2013toa, PhysRevD.78.063506, Delgado:2020htr}.  

\subsection{Asymmetric bounce}

In reference \cite{Delgado:2020htr} it is shown that, within the dBB interpretation, an initial state of the form 
\begin{equation}\label{non unitary initial wf}
   \Psi_{0}(\chi)=\left( \frac{8}{\pi\sigma^{2}}\right)^{\frac{1}{4}} 
    \exp \left( -\frac{\chi^2}{\sigma^2} +\imathnew p \chi    \right)\,,
\end{equation}
with $p,\sigma \in \mathds{R}$, can be evolved by the propagator 
\begin{equation}\label{nu propagator}
    G^{NU}(\chi,\chi_{0},T)=\sqrt{-\frac{\imathnew}{\pi T}} \exp \left[-\frac{\imathnew (\chi-\chi_{0})^{2}}{T}\right]\,,
\end{equation}
yielding the following solution for all times
\begin{align}\label{Psi_evolved_t}
\Psi(\chi,T)=&R_\Psi(\chi,T)\exp{[\imathnew S_\Psi(\chi,T)]}\,,\\
\nonumber    R_\Psi(\chi,T)\equiv&\left[\frac{8\sigma^2}{\pi(\sigma^4+T^2)}\right]^{1/4}\times\\
\nonumber &\times \exp\left[-\frac{\sigma^2}{\sigma^4+T^2}\left(\chi+\frac{pT}{2}\right)^2\right]\,,\\
\nonumber    S_\Psi(\chi,T)\equiv& p\left(\chi+\frac{pT}{4}\right)-\frac{T}{\sigma^4+T^2}\left(\chi+\frac{pT}{2}\right)^2+\\
 \nonumber   &+\frac{1}{2}\arctan\left(\frac{T}{\sigma^2}\right)\,.
\end{align}

The quantum parameters $\sigma$ and $p$ are related to the initial width (with a standard deviation $\sigma/\sqrt{2}$) and momentum of the Gaussian in configuration space, respectively. The associated Bohmian trajectories for the variable $\chi$, which can be translated to the trajectories of the scale factor $a$ through definition \eqref{chi and a}, are a solution to the Bohmian guidance equation
\be
\frac{\dif\chi}{\dif T}=-\left.\frac{1}{2}\frac{\partial S_\Psi}{\partial \chi}\right\rvert_{\chi(T)}=\frac{T}{\sigma^4+T^2}\chi-\frac{p\sigma^4}{2(\sigma^4+T^2)}\,,
\label{guidance_Bohm}
\ee
which can be integrated to give
\begin{equation}
    \chi(T) = \chi_{b} \biggl[ 1+ \left( \frac{T}{\sigma^{2}} \right) ^{2}+ \left( \frac{p}{2 \chi_{b}} \right) ^{2}\left(\sigma^{4}+T^{2}\right) \biggr]^{\frac{1}{2}} - \frac{pT}{2}\,,
    \label{chisol}
\end{equation}
where $\chi_{b}$ is the value of the variable $\chi$ at the bounce, occurring at $T_{b}=p\sigma^{4}/(2 \chi_{b})$.

Defining 
\begin{equation}\label{y_expression}
    \bar{T}\equiv \frac{T}{\sigma^2}\,, \quad
    x_b\equiv \frac{a_0}{a_b}\,,\quad
    y\equiv-\frac{p\sigma^2}{2\chi_b}\,,
\end{equation}

we can rewrite \eqref{chisol} in the simple form
$$\chi(\bar{T})=\chi_b\left(y \bar{T}+\sqrt{1+y^2}\sqrt{1+\bar{T}^2}\right)\,.$$
As a result, the scale factor reads 
\begin{equation}\label{scale_factor}
    a(\bar{T})=a_b \left(y \bar{T}+\sqrt{1+y^2}\sqrt{1+\bar{T}^2}\right)^{\frac{2}{3(1-w)}}\,,
\end{equation}
where $a_b$ is the scale factor at the bounce, and $\bar{T}_b=-y$ corresponds to the time at which the bounce occurs. When $|T|\gg \sigma^2$, or $|\bar{T}|\gg1$, we have $a(\bar{T})\propto |\bar{T}|^{\frac{2}{3(1-w)}}$ which, translated to cosmic time, corresponds to a classical single-fluid Friedmann evolution $a(\tau)\propto |\tau|^{\frac{2}{3(1+w)}}$.

Looking at Eq.~\eqref{scale_factor}, we can see that the time asymmetry of the model comes from the linear term in $\bar{T}$, with the property $a(-y,\bar{T})=a(y,-\bar{T})$. Hence, changing $y\rightarrow -y$ is equivalent to time-reversing the original solution.
From now on we will choose $y>0$ (or $p<0$), the case $y<0$ being straightforwardly obtained by time reversing the conclusions.

The Hubble function is given by $H\equiv \dot{a}/(Na)=a^{-3(1+w)/2}\left(\dif\chi/\dif T\right)$. Its expression squared reads

\begin{equation}\label{Hubble function}
H^2(\bar{T})=\frac{4}{9(1-w)^2}\frac{a_b^{3(1-w)}}{\sigma^4 a^{3(1+w)}(\bar{T})} F(\bar{T})\,,
\end{equation}
where
\begin{equation}\label{F}
F(\bar{T})=\left(y+\sqrt{1+y^2}\frac{\bar{T}}{\sqrt{1+\bar{T}^2}}\right)^2\,.
\end{equation}
In the asymptotic limits we get

\begin{equation}
   \lim_{\bar{T}\rightarrow\pm\infty} F(\bar{T})=:F_{\pm}=\left(y\pm\sqrt{1+y^2}\right)^2\,,
   \label{asympF}
\end{equation}
which is a constant. Thus, we get
\begin{equation}
H^2_{\pm}(\bar{T})\propto\frac{F_{\pm}}{a_\pm^{3(1+w)}(\bar{T})}\propto\rho_{\pm}(\bar{T})\,,
\end{equation}
which are the asymptotic classical Friedmann equations at both limits, as expected (the subscripts $\pm$ refer to the asymptotic future and past, respectively). 

Fixing the same scale factor at both asymptotic classical phases, we have 
\begin{equation}
H^2_{\pm}(a)\propto\frac{F_{\pm}}{a^{3(1+w)}}\propto\rho_{\pm}(a)\,,
\label{H23}
\end{equation}
showing that the model is not symmetric. In fact, the conserved quantities $F_{\pm}$, which can be understood as the total number of particles inside a volume cell for a given scale factor in both asymptotic phases, are different, see Eq.~\eqref{asympF}, with creation of particles from the asymptotic past to the asymptotic future: 

\begin{equation}
\frac{\rho_+(a) a^{3(1+w)}}{\rho_-(a) a^{3(1+w)}} = \frac{F_+}{F_-}=\frac{\left(y+\sqrt{1+y^2}\right)^2}{\left(y-\sqrt{1+y^2}\right)^2} > 1 \,.
\label{ratio}
\end{equation}
For $y\gg 1$, this growth can be huge: from \eqref{H23} and \eqref{ratio}, evaluating at the same scale factor, we have that the respective ratio of the energy densities at the asymptotic future and past is given by

\begin{equation}
\frac{\rho_+(a) a^{3(1+w)}}{\rho_-(a) a^{3(1+w)}}=\frac{\rho_+(a)}{\rho_-(a)} =\frac{H_+^2(a)}{H_-^2(a)} \approx 16 y^4 \gg 1 \,,
\label{ratio2}
\end{equation}
implying both a large creation of particles, and a very slow asymptotic contraction when compared to the asymptotic expansion rate for the same $a$, $H_+ \gg |H_-|$. The ratio of Hubble radii $R_\pm$  then reads

\begin{equation}
\frac{R_-(a)}{R_+(a)} =\frac{H_+(a)}{|H_-(a)|} \approx 4 y^2 \gg 1 \,.
\label{ratio3}
\end{equation}
The classical scale factor in both asymptotic limits simplifies to
\begin{eqnarray}
    a_{\mathrm{expansion}}&\approx& a_b[(2y)\bar{T}]^{\frac{2}{3(1-w)}}\,,\\
    a_{\mathrm{contraction}}&\approx& a_b[|\bar{T}|/(2y)]^{\frac{2}{3(1-w)}}\,.
\end{eqnarray}

Comparing $H^2_{+}(\bar{T})$ with the late time Friedmann equation $H^2/H_0^2=\Omega_{w,0}x^{3(1+w)}$, where $x\equiv a/a_b$ and $\Omega_{w,0}$ is the ratio between the fluid energy density and the critical density when $H=H_0$, we can relate $\Omega_{w,0}$ and the Hubble radius today $R_{H_0}\equiv 1/H_0$ to the free parameters of the model: 

\begin{equation}
\sigma^2 a_b^{3w} = \frac{2 R_{H_0} \left(y+\sqrt{1+y^2}\right)}{3(1-w) x_b^{3(1+w)/2} \sqrt{\Omega_{w,0}}} \,,
\label{parameters}
\end{equation}
which will be very useful in the next section.

\subsection{Quasi de Sitter phase}
\label{secqdS}

\begin{figure}
    \centering
    \includegraphics[scale=0.55]{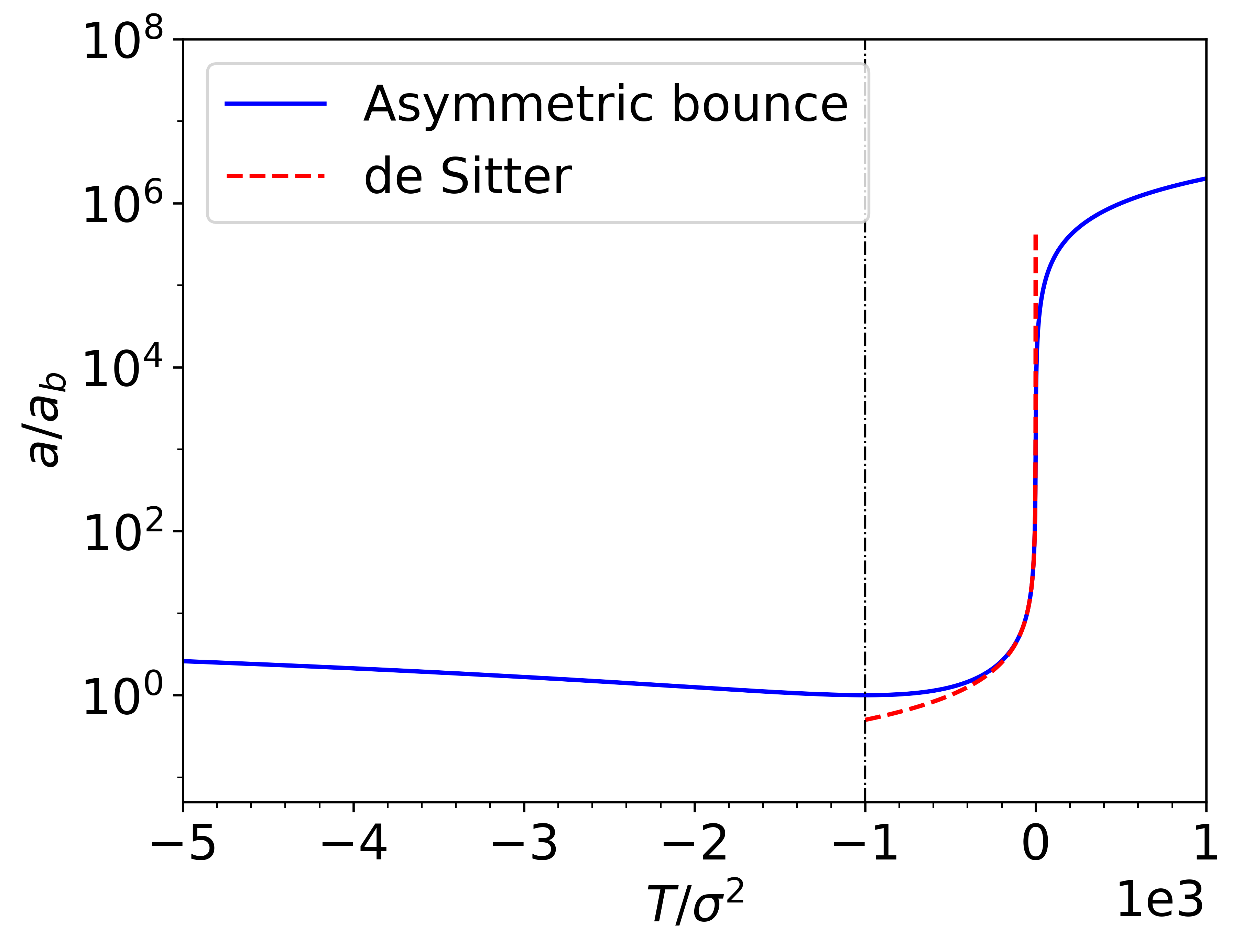}
    \caption{Evolution of the scale factor $a$ in the scenario with a subtle contraction, compared to the de Sitter evolution $a\propto \eta^{-1}$. We use $y=10^3$ and $w=1/3$, corresponding to a radiation fluid. The vertical line indicates the bounce time $\bar{T}_b=-y$. }
    \label{bounce}
    \end{figure}

A noteworthy feature in the scenario with $y\gg 1$ arises from the fact that for $\bar{T}<0$, $|\bar{T}|\gg1$ and $y\gg1$ the scale factor \eqref{scale_factor} reads
\begin{equation}
    a(\bar{T})\approx a_b\left( \frac{|\bar{T}|}{2y}+\frac{y}{2|\bar{T}|}\right)^{\frac{2}{3(1-w)}}\,.
    \label{aapprox}
\end{equation}
The classical contraction is recovered when the first term in \eqref{aapprox} dominates, i.e. when $|\bar{T}|\gg y$. 
In this case, $a\propto |\eta|^{\frac{2}{1+3w}}$, as $|\eta|\propto |\bar{T}|^{\frac{1+3w}{3(1-w)}}$. On the other hand, when $|\bar{T}|\ll y$, which happens after the bounce, the second term in \eqref{aapprox} dominates and, since in this case $|\eta|\propto |\bar{T}|^{\frac{5-9w}{3(1-w)}}$, we have
\begin{equation}
    a(\eta)\propto |\eta|^{-\frac{2}{5-9w}}\,.
    \label{adesitter}
\end{equation}
Hence, an intermediate phase between the bounce and the classical expansion arises naturally, corresponding to a quantum accelerated expansion when $1/3\leqslant w < 5/9$\footnote{Note that for $w<1/3$, which includes the matter bounce scenario, one gets an intermediate phantom accelerated expansion. This is an interesting possibility, which might be explored in future investigations.}.
If $w=1/3$, i.e. in a radiation-dominated universe, this phase is in fact a de Sitter expansion, where $a(\eta)\propto \eta^{-1}$. Moreover, since \eqref{adesitter} comes from an approximation, the quantum accelerated expansion is indeed a quasi-de Sitter (qdS) phase. Accordingly, we find the Hubble parameter around this stage to be almost - but not exactly - constant. The corresponding scale factor is depicted in Figure \ref{bounce}, where a comparison between the quantum accelerated phase and the de Sitter scale factor is exhibited.

Since $y\gg 1$, this quantum accelerated expansion might last for a long time, around $-y\ll \bar{T}\ll -1$. Afterwards, when $\bar{T}\gg 1$, the classical decelerated expansion of the standard cosmological model takes place. In other words, the scenario determined by the condition $y\gg 1$ and $w\approx 1/3$ encompasses a bounce regulated by quantum effects, followed by a large period of accelerated quantum expansion, which gracefully exits into the usual classical regime. Therefore, this scenario is analogous to inflation without an inflaton and without the usual reheating phase, as it is smoothly followed by the classical decelerated expansion dominated by radiation. As discussed after Eq.~\eqref{ratio2}, there is a huge creation of photons during the quantum phase.

The effective energy density of this inflationary period can be obtained by expanding the Hubble parameter for large $y$ and then large $\bar{T}$. The dominant constant term, present only in this phase, is found to be equal to $\Omega_{\mathrm{qdS}}=(x_b/y)^4\Omega_{w,0}$. We shall later express this in a more suggestive way.\\

The scale factor \eqref{scale_factor} contains three parameters: two from the initial Gaussian wave function ($\sigma,p$), and one as an integration constant ($a_b$ or, equivalently, $x_b$) of the guidance equation \eqref{guidance_Bohm}. They are tied together by the late time Hubble parameter expansion, see Eq.~\eqref{parameters}, leaving the set $\{y,x_b\}$ as free parameters. A physically important parameter is the minimum curvature scale $l_c$ of the model, which cannot be very close to the Planck scale, where the above quantization scheme is not reliable. It is given by 
\be
l_c\equiv\frac{1}{\sqrt{\max[R(\bar{T})]}}\,,
\label{max_curvature_scale}
\ee
where $R(\bar{T})=6\left(\Ddot{a}/a+H^2\right)$ is the Ricci scalar computed from the background trajectory (here an overdot denotes a cosmic time derivative). It reads
\begin{align}
   R=&\frac{4\alpha^{-2\left(\frac{1+w}{1-w}\right)}}{(1-w)\left(\sigma^2a_b^{3w}\right)^2}\times \nonumber \\
    &\times\left[\frac{\sqrt{1+y^2}}{(1+\bar{T}^2)^{3/2}}\alpha+\frac{1-3w}{3(1-w)}\left(\frac{\dif \alpha}{\dif \bar{T}}\right)^2\right]\,,
    \label{Ricciexact}
\end{align}

where $\alpha\equiv y\bar{T}+\sqrt{1+y^2}\sqrt{1+\bar{T}^2}$. 

In the case of interest, i.e. $w=1/3$, which leads to the qdS expansion, and large asymmetry ($y\gg 1$), the maximum curvature happens when \cite{Delgado:2020htr}
\begin{eqnarray}
    \bar{T}_{\mathrm{max}}=-\sqrt{\left(\sqrt{1+y^2}-1\right)/2}\approx-\sqrt{y/2}\,.
\end{eqnarray}

Hence, from Eqs.~\eqref{parameters} and \eqref{Ricciexact}, the minimum curvature scale in terms of the previous parameters and cosmological observables reads
\be
l_c\equiv\frac{y^2}{x_b^2\sqrt{12\Omega_{r,0}}}R_{H_0}\,,
\label{lc_expression}
\ee
where $\Omega_{r,0}\approx\num{9e-5}$ is the density parameter of radiation today. The curvature scale is bounded from below so as not to be too close to the Planck scale, $l_P\approx 10^{-61}R_{H_0}$, at an energy scale of about $10^{19}\,\mathrm{GeV}$. Both the minisuperspace approximation and the Wheeler-DeWitt quantization cease to be valid or meaningful close to this boundary, requiring a full, yet unknown, theory of quantum gravity. On the other hand, the model must recover its standard FLRW evolution much before nucleosynthesis, in order not to spoil its good agreement with observations, thus leading to an upper bound for $l_c$.

Using Eqs.~\eqref{scale_factor} and \eqref{Hubble function} for $w=1/3$, we can estimate the Hubble parameter in the classical limit around nucleosynthesis:
\be
H\approx \frac{1}{\sigma^2a_b(2y\bar{T}^2)}=\frac{1}{8\sqrt{3}l_c \bar{T}^2}\,,
\ee
This approximation is good from $\bar{T}\approx10$ onward, where the relative error with respect to classic evolution is about $0.5\%$. From the characteristic energy density at the dawn of the nucleosynthesis era, $\varepsilon_{\mathrm{nucleo}}\approx 10^{16}\,\mathrm{g}\,\mathrm{cm}^{-3}$, and using the Friedmann equation $H^2=H_0^2\Omega(\bar{T})$ we get
\be
H^2/H_0^2>\Omega_{\mathrm{nucleo}}\to \frac{l_c}{R_{H_0}}<\frac{1}{8\sqrt{3}(10)^2\sqrt{\Omega_{\mathrm{nucleo}}}}\,.
\ee
Thus, demanding that the transition of the quantum background to the classical behavior must take place much before nucleosynthesis, and does not reach the Planck scale by three orders of magnitude, we find the constraint
\be
10^{3}<\frac{l_c}{l_P}\ll 10^{35}\,.
\label{lc_bounds}
\ee

Having introduced $l_c$, the effective qdS energy density may be rewritten as $\Omega_{\mathrm{qdS}}=(R_{H_0}/l_c)^2/12$, which in turn gives an effective primordial cosmological constant directly related to the minimum curvature scale,
\be
\Lambda_{\mathrm{qdS}}\equiv \frac{1}{4l_c^2}\,,
\label{effective_Lambda_qdS}
\ee
being responsible for driving the almost exponential expansion. Note this can also be obtained by comparing $a\approx y/(2|\bar{T}|)$ with the de Sitter scale factor $a=(H|\eta|)^{-1}$, with $H=\sqrt{\Lambda/3}$.

Furthermore, an initial estimate of the $\mathrm{e}-$folds of this model can be made through
\be
\mathcal{N}=\ln\left(\frac{a_f}{a_i}\right)\approx \ln\left[ \frac{a(0)}{a(-y)}\right]\approx \ln y \,,
\label{e-folds}
\ee
where we considered the beginning of the inflationary period to be the bounce and its end to be at $\bar{T}=0$ \footnote{Since there is a very slow contraction, the initial time for $a_i$ is very close to the start of the qdS phase; similarly, the transition to the classical expansion occurs around $\bar{T}\sim\mathcal{O}(1)$, and does not change the estimate significantly.}. 

To accommodate for e.g. $\mathcal{N} > 60$ we should require
\be
 y  > \num{1.14 e26} \,.
 \label{e-folds bounds}
\ee

\section{COSMOLOGICAL PERTURBATIONS: ANALYTICAL RESULTS}
\label{secanalytical}

The variable generally used to describe the evolution of scalar perturbations on a homogeneous and isotropic spacetime with a single perfect fluid is the gauge invariant curvature perturbation $\zeta$, which is a combination of fluid and (scalar) metric linear perturbations \cite{PhysRevD.76.023506}. The apropriate variable to be quantized is the Mukhanov-Sasaki variable, which in the perfect fluid case is related to the curvature perturbation $\zeta$ in momentum space through $v_k\equiv a\zeta_k$. It satisfies the equation of motion
\begin{equation}
    v_k''+\left(c_s^2 k^2-\frac{{a}''}{a}\right)v_k=0\,,
    \label{mukhanov-sasaki1}
\end{equation}
where a prime denotes derivative with respect to conformal time.

In the subsections below, we will present the analytical results for arbitrary $w$, and then particularize to $w=1/3$.

\subsection{Arbitrary perfect fluids: general $w$}
\label{general w}

Let us begin by analyzing the limiting cases of \eqref{mukhanov-sasaki1} and matching them at the crossing $c_s^2k^2\approx a''/a$ \footnote{Note that the classical FLRW Ricci scalar may be written as $R=6 a''/a^3$ (for the case of exact radiation it vanishes identically). Moreover, for the (classical) pure de Sitter or perfect fluid cases, $R\propto H^2$. The physical curvature scale $R^{-1/2}$ may be written as $a\ell$, where $\ell\propto\sqrt{a/a''}$ is the comoving curvature scale. Therefore, apart from a factor of order unity, the potential crossing condition $c_s^2k^2=a''/a$ can also be understood as the time when the physical scale $a/k$ becomes equal to the curvature scale, sometimes called Hubble crossing.}. When $c_s^2k^2\gg {a}''/a$, the solution is given by
\begin{equation}
    v_k\approx C_1(k)\mathrm{e}^{\imathnew c_sk\eta}+C_2(k)\mathrm{e}^{-\imathnew c_sk\eta}\,,
\end{equation} 
while for $c_s^2k^2\ll {a}''/a$ the solution can be approximated as
\begin{equation}
    v_k\approx  A_1(k)a+A_2(k)a\int\frac{\dif \eta}{a^2}+\mathcal{O}(k^2)\,.
    \label{vk_super_Hubble}
\end{equation}

For the perturbation modes that cross the potential $V\equiv a''/a$ during the contracting phase, the $A_2(k)$ term - which grows when $a$ decreases - dominates in the decelerated expansion. 

On the other hand, for modes that cross the potential during the quantum accelerated expansion after the bounce, it is $A_1(k)$ that dominates in the decelerated expansion. 

In order to find $A_1(k)$ and $A_2(k)$, we match both approximate solutions at the crossing $c_s^2k^2\cong V$. Since for $\bar{T}\rightarrow -\infty$, ${a}''/a \rightarrow 0$, we can choose the initial condition as the normalized adiabatic vacuum $v_k\approx \exp{(-\imathnew c_sk\eta)}/\sqrt{2c_s k}$. Far from the bounce into the remote past, $\bar{T}\ll-y$, we can approximate the potential ${a}''/a$ using the first term in \eqref{aapprox}, yielding
\begin{equation}
    V \propto \left(\frac{1}{|\bar{T}|}\right)^{\frac{2(1+3w)}{3(1-w)}}\,.
\end{equation}
This phase corresponds to a classical contraction with $a(\eta)\approx|\eta|^{\frac{2}{1+3w}}$, $|\eta|\propto |\bar{T}|^{\frac{1+3w}{3(1-w)}}$. Since the term that dominates the potential is proportional to $|\eta|^{-2}$, we have the horizon crossing at conformal time $\eta_c$ when $c_s^2k^2\approx \eta_c^{-2}$. From the approximation \eqref{vk_super_Hubble} we then have 
\begin{equation}
    v_k=A_1(k) \eta^{\frac{2}{1+3w}}+c_1 A_2(k)\eta^{\frac{-1+3w}{1+3w}}\,, 
    \label{tomatch}
\end{equation}
where $c_1$ is a constant of order unity. Matching the adiabatic vacuum and \eqref{tomatch} at $\eta_c\approx k^{-1}$ for both $v_k$ and ${v_k}'$ we find
\begin{align}
    A_1(k)&\propto k^{\frac{3(1-w)}{2(1+3w)}}\,,\\
    A_2(k)&\propto k^{-\frac{3(1-w)}{2(1+3w)}}\,.
\end{align}

As $A_2(k)$ dominates for modes that cross the potential during the decelerated contraction, the scalar power spectrum reads
\begin{equation}
    P_\zeta(k)\propto k^3|A_2(k)|^2\propto k^{\frac{12w}{1+3w}}\,,
    \label{power_spectrum_contracting}
\end{equation}
see Ref.~\cite{PhysRevD.78.063506} for details.

The almost scale invariant behavior, compatible with observations, is then obtained for $w\approx 0$. This corresponds to the result in \cite{PhysRevD.78.063506} for the quantum symmetric bounce, to which the present asymmetric case reduces as $y\to0$ \footnote{It is worth noting that the main $|\eta|^{-2}$ contribution to the potential in the contracting phase vanishes for $w=1/3$. One finds that the next contribution corresponds to a $|\eta|^{-\frac{8}{1+3w}}$ behavior, thus leading to $\left.V\right\rvert_{w=1/3}\propto |\eta|^{-4}$. This is typical of an evolution which is almost, but not exactly, classically dominated by radiation. Proceeding in the same way as before, the dominant term $A_2(k)$ goes as $k^{-1/2}$, leading to a blue tilted power spectrum $P_{\zeta,1/3}(k)\propto k^2$, which coincides with \eqref{power_spectrum_contracting} evaluated for radiation.}. 

Let us now investigate the quantum accelerated phase, considering the potential obtained from the second term in \eqref{aapprox}. For $-y\ll \bar{T}\ll -1$, we have \eqref{adesitter}, and the potential is approximately
\begin{equation}
  V \propto \left(\frac{1}{|\bar{T}|}\right)^{\frac{2(5-9w)}{3(1-w)}} \propto |\eta|^{-2}\,. 
    \label{V_expansion_qdS}
\end{equation}
Consequently, $\eta_c\approx k^{-1}$ also for this phase. Using \eqref{vk_super_Hubble} we find 
\begin{equation}
    v_k\approx A_1(k)\eta^{-\frac{2}{5-9w}}+c_2A_2(k)\eta^{\frac{7-9w}{5-9w}}\,.
\end{equation}
Matching both $v_k$ and ${v_k}'$ with the adiabatic vacuum at $\eta_c\approx k^{-1}$, we find
\begin{align}
    &A_1(k)\propto k^{-\frac{9(1-w)}{2(5-9w)}}\,,\\
    &A_2(k)\propto k^{\frac{9(1-w)}{2(5-9w)}}\,.
\end{align}
The quantum accelerated expansion is dominated by $A_1(k)$ and, therefore, the scalar power spectrum has the following $k-$dependence:
\begin{equation}
    P_\zeta(k)\propto k^3|A_1(k)|^2\propto k^{\frac{6(1-3w)}{5-9w}}\,.
    \label{Pzeta_w_dep}
\end{equation}

Note that almost scale invariance is attained for $w\approx 1/3$, as expected from the qdS behavior for this choice of fluid. Thus, if the background is dominated by a radiation fluid and the cosmological scales that we observe today cross the potential during the quantum accelerated expanding phase, then their power spectrum will be almost scale invariant. In this case, the accelerated period is a qdS expansion akin to inflation, with the advantage of happening naturally within the model, without an inflaton. Let us then focus our attention to the radiation fluid, $w=1/3$.

\subsection{The case of radiation: $w=1/3$}

In this case, the scale factor has the simple form

\begin{equation}
    a(\bar{\eta}) = a_b \left( y\bar{\eta} + \sqrt{1 + y^2}\sqrt{1 + \bar{\eta}^2} \right) \,,
    \label{eq2:aw=1/3}
\end{equation}
and the bounce happens when $\bar{\eta}=-y$. Recall that for $w=1/3$ the time $\bar{T}$ is a dimensionless conformal time, which we have named $\bar{\eta}$. Equation~\eqref{mukhanov-sasaki1} reads

\begin{equation}
    v_k''+\left(\bar{k}^2-\frac{{a}''}{a}\right)v_k=0\,,
    \label{mukhanov-sasaki1.5}
\end{equation}
where the primes now denotes derivatives with respect to $\bar{\eta}$, $\bar{k}\equiv \sigma^2 c_s k$, and 
\be
V\equiv \frac{{a}''}{a}=\frac{\sqrt{1+y^2}\left(1+\bar{\eta}^2\right)^{-3/2}}{ y \bar{\eta}+\sqrt{1+y^2}\sqrt{1+\bar{\eta}^2}}\,.
\label{potential_exact}
\ee
We will assume from now on that $y\gg1$. 

Let us make a brief qualitative summary of the history of the background and perturbations. The periods (1-3) in what follows are for $\bar{\eta}<0$, the period (4) is the transition from $\bar{\eta}$ negative to $\bar{\eta}$ positive, and period (5) for $\bar{\eta}>0$.
\vspace{0.3cm}

1) For $-\bar{\eta}\gg y$:

In the far past of the model, the universe is contracting from $\bar{\eta} \rightarrow -\infty$ as $a \approx -a_b \bar{\eta}/(2 y)$, a classical contraction dominated by radiation. The potential $V\equiv a''/a$, also called the effective Hubble parameter, goes as $2 y^2/\bar{\eta}^4$. Note that $V$ is not zero because $a$ is not exactly $-a_b \bar{\eta}/(2 y)$. This is typical of an evolution which is almost, but not exactly, classically dominated by radiation.

As we have seen in the previous subsection, the wavenumbers that cross the effective Hubble parameter at this epoch will not be scale invariant, with an associated power spectrum scaling as $\bar{k}^2$, so we must guarantee that they are very small (very large scales, much bigger than the Hubble radius today).
\vspace{0.3cm}

2) For $-\bar{\eta}\approx y$:

In this phase the quantum effects become important, they realize the bounce and launch the universe in a quantum expanding phase. 
\vspace{0.3cm}

3) For $1 \ll -\bar{\eta}<y$:

The universe enters in a expanding phase with $a \approx -a_b y/(2 \bar{\eta})$, which is typical of a de Sitter expansion. It is a quantum effect (note that the background fluid is always radiation). The potential $V\equiv a''/a$ goes as $2/\bar{\eta}^2$.

The wavenumbers that cross the effective Hubble parameter at this epoch will be almost scale invariant because they cross the potential in a qdS phase, as in inflation\footnote{Note that for $w=1/3$ exactly the spectral index is not sufficiently red-tilted, see Fig.~\ref{figpknew} below. Only for $w= 1/3+\epsilon$, $\epsilon\approx 3.83 \times 10^{-3}$, one can get $n_s\approx 0.96$. In the Conclusion we will return to a discussion about this point.}. Hence, we have a bounce naturally followed by an inflationary phase. The cosmological large scales observed in the Planck satellite should cross the effective Hubble parameter in this epoch. 

\vspace{0.3cm}

4) For  $-\mathcal{O}(1) < \bar{\eta} < \mathcal{O}(1)$:

In this phase the maximum of the potential, or of the effective Hubble parameter, is reached. With $y\ll 1$, it happens for $\bar{\eta}=-\sqrt{2}/4$, and the maximum value of the effective Hubble parameter is $V=32/27\approx 1.18$. Wavenumbers larger than this value will never cross the effective Hubble parameter, hence they behave typically as the ultraviolet limit of the Minkowski vacuum. This means that there is a cutoff for the perturbation modes, beyond which no crossing occurs. The comoving wavenumbers which are bigger than the maximum effective Hubble parameter, $\bar{k}^2 > V_{\rm max} = 1.18$, never feel the evolution of the universe. Therefore, $\bar{k}_{\rm max} = \sqrt{1.18}=1.09$ is our cutoff scale. The physical scales of these wavenumbers will be evaluated below.
\vspace{0.3cm}

5) For $\bar{\eta} \gg 1$:

In this era we have $a \approx a_b (2 y \bar{\eta})$, and the potential is such that $V\propto 1/\bar{\eta}^4$. The classical radiation-dominated expanding phase is recovered. Therefore, we have a natural graceful exit from inflation to the standard model radiation-dominated era.

\begin{figure}
    \centering
    \includegraphics[scale=0.55]{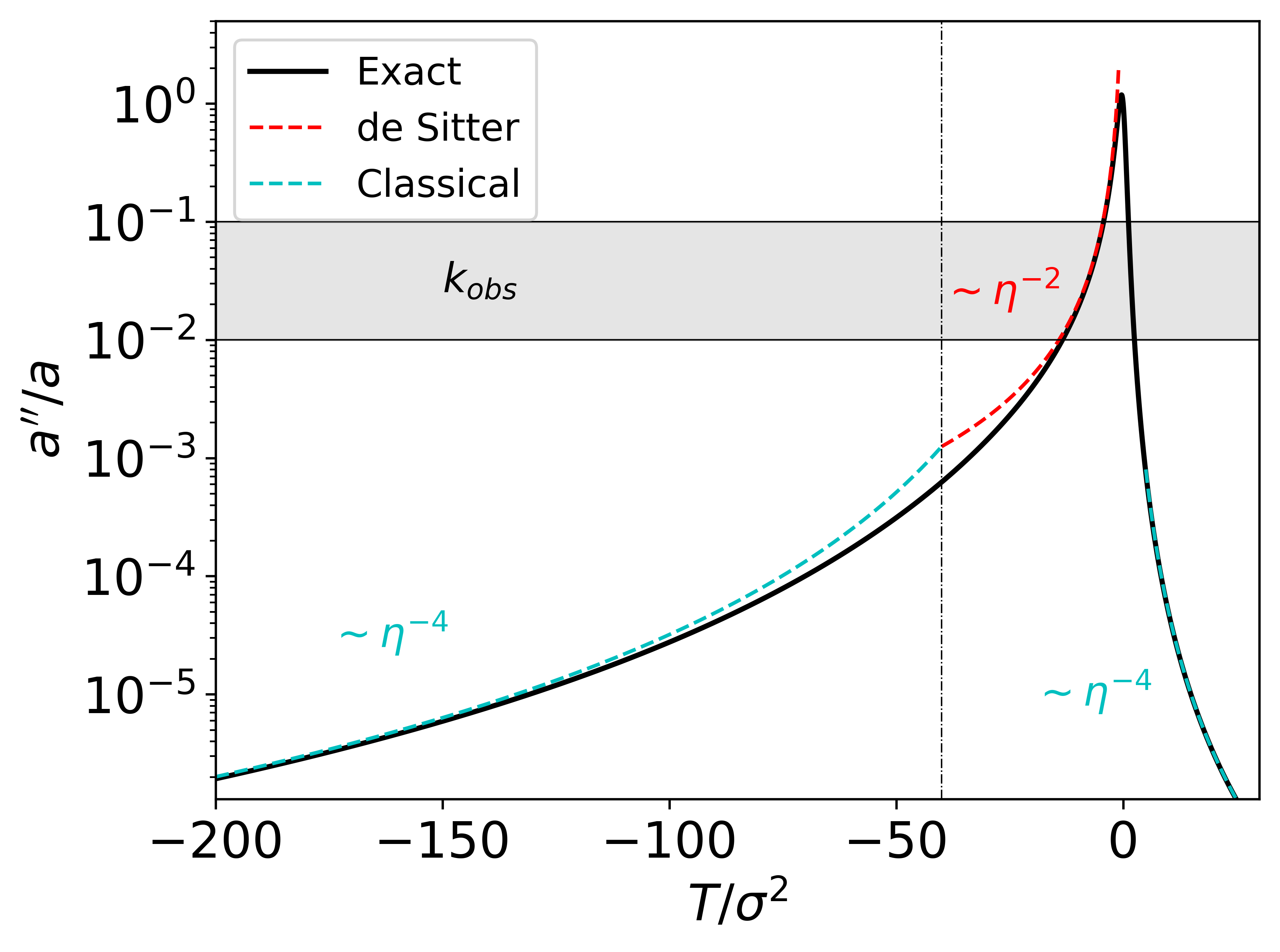}
    \caption{Perturbation potential $a''/a$ for the radiation case and its limiting regimes for $y=40$: in green, $V$ during the classical contraction and expansion; in red, $V$ during the qdS phase. The vertical line represents the bounce, while the shaded area depicts the range of observable co-moving wave-numbers $k$, which constrains the parameters of the model.}
    \label{potential_obsscales}
\end{figure}

Figure \ref{potential_obsscales} shows the potential $V$ given in Eq. \eqref{potential_exact} for a representative value of $y$, highlighting its approximate regimes discussed above. The crossing condition $\bar{k}=\sqrt{V}$ for a given comoving wavenumber within the observable range is set to take place in the accelerated phase. As mentioned above, one can see the maximum value of the potential for $y\gg 1$ is given by 
\be
V(\bar{\eta}_{\mathrm{max}})\approx\frac{32}{27}\approx 1.18 \Rightarrow \bar{k}_{\rm max}\approx 1.09 \,,
\label{potential_cutoff}
\ee
at $\bar{\eta}_{\mathrm{max}}\approx-\frac{\sqrt{2}}{4}$. Finally, we point again that the time parameter in Fig.~\ref{potential_obsscales} and subsequent ones is, for $w=1/3$, $T/\sigma^2=\eta/\sigma^2=\bar{\eta}$. 

\subsubsection{\textbf{Relations between the free parameters and physical scales}}

Let us define 
$${k_H}\equiv k R_{H_0}/a_0=R_{H_0}/\lambda_{\rm{ physical},0}\,,$$
which is the ratio of the Hubble radius today to the physical wavelength of the mode $k$ today. Hence, the large cosmological wavelengths observed in the CMB are in the approximate range $1<{k_H}<10^3$ \cite{Planck:2018jri}, which corresponds approximately to $10\; {\rm Mpc}<\lambda_{\rm {physical},0}<10\; {\rm Gpc}$ today.

The relationship between $k_H$ and $\bar{k}$ reads

\begin{equation}
{k_H} =\frac{R_{H_0}}{c_s(\sigma^2 a_b)x_b}\bar{k}\,.
\label{ktil}
\end{equation}
Using Eqs.~\eqref{parameters} and \eqref{lc_expression} for $w=1/3$ and $y\gg 1$ we get

\begin{equation}
{k_H} =\sqrt{\frac{R_{H_0}}{l_c}}
\left(\frac{\Omega_{r,0}}{12}\right)^{1/4}\frac{\bar{k}}{2c_s}\equiv C \bar{k}\,.
\label{ktil2}
\end{equation}

Note that $C$ is a very large constant, so that the cosmological wavenumbers are associated with very small $\bar{k}$. We impose that these cosmological wavenumbers cross the effective Hubble parameter in the quantum inflationary phase, namely, when $1\ll -\bar{\eta}<y$, yielding ${\bar{k}}_{\rm cross}=1/{|\bar{\eta}}_{\rm cross}|$, which makes ${\bar{\eta}}_{\rm cross}$ very big. This is another reason to take $y\gg 1$, as this crossing must happen when $-\bar{\eta}<y$. 

Knowing that $R_{H_0}/l_P\approx 10^{61}$, and from the range for $l_c$ given in Eq.~\eqref{lc_bounds}, we get

\begin{equation}
10^{15}\ll C < 10^{28}\,,
\label{range22}
\end{equation}
and $y>C$.

These huge values of $y$ may be frightening, but in fact they are good because the effective Hubble parameter (the potential) becomes very closely independent of $y$ in this regime,

\begin{equation}
V\approx \frac{1}{(1+\bar{\eta}^2 )^{3/2}(\bar{\eta} + \sqrt{1+\bar{\eta}^2})}\,.
\label{Vap}
\end{equation}
This approximation is excellent for any $\bar{\eta}>-y$, but it is not good otherwise. However, as we are interested in wavenumbers that cross the effective Hubble parameter only when the quantum inflationary expansion is under way - which takes place after the bounce - we can say that, for a period satisfying $-\bar{\eta}<y$, these modes satisfy $\bar{k}^2 \gg V$. Hence, we can pose vacuum initial conditions there, and all the numerical calculations are made when Eq.~(\ref{Vap}) is an excellent approximation.

\subsubsection{\textbf{Analytical estimates of the power spectrum}}

The curvature perturbation $\zeta_k$ satisfies the general equation in arbitrary time:

\begin{equation}
\ddot{\zeta}_k + \frac{\dot{m}}{m}\dot{\zeta}_k + \nu^2 \zeta_k  = 0 \,,
\label{eq:zeta}
\end{equation}
where $m$ and $\nu$ are the generalized time-dependent mass and frequency. All General Relativity (GR) linear scalar perturbations with one field as the source term can be put in this form. The generalized Mukhanov-Sasaki variable is $v_k=\sqrt{m}\,\zeta_k$, and satisfies,

\begin{equation}
\ddot{v}_k + \left(\nu^2-\frac{\ddot{\sqrt{m}}}{\sqrt{m}}\right) v_k = 0 \,,
\label{eq:MS0}
\end{equation}

For perfect fluids, $m$ and $\nu$ can be read from the GR action as (see Eqs.~(92-94) of Ref.~\cite{Vitenti:PhysRevD.87.103503}),

\begin{equation}
m = \frac{(\rho+p)a^3}{N c_s^2 H^2} \,,\quad \nu=\frac{N k c_s}{a} \,,
\label{mnu}
\end{equation}
where $N$ is the lapse function and $p=w\rho$.

The vacuum (or WKB) initial condition for $\zeta_k$ can be set in the epoch when $\nu^2\gg \ddot{\sqrt{m}}/\sqrt{m}$, yielding $|v_k^{\rm WKB}|=1/\sqrt{2\nu}$. In our case, this phase includes the period of classical evolution, where we can use the Friedmann equation in the contracting branch of the evolution. We therefore obtain that

\begin{equation}
|\zeta_k^{\rm WKB}| =\frac{|v_k^{\rm WKB}|}{\sqrt{m}} = \frac{1}{\sqrt{2m\nu}}= \frac{l_P}{a} \sqrt{\frac{4 \pi c_s}{3(1+w)k}}\,,
\label{WKBz}
\end{equation}
where in our convention $G=l_P^2$.

On the other hand, we will work with the equation with dimensionless variables and parameters,

\begin{equation}
    v_{k,(2)}^{\prime \prime} + \left(\bar{k}^2 - \frac{A^{\prime \prime}}{A} \right) v_{k,(2)} = 0\,,
    \label{eq:MSn}
\end{equation}
where we take 
$$ A\equiv y\bar{\eta} + \sqrt{1 + y^2}\sqrt{1 + \bar{\eta}^2} \,, $$
as $a_b$ disappears from Eq.~(\ref{eq:MSn}), and 
$$\zeta_{k,(2)}=v_{k,(2)}/A\,.$$ 
The WKB curvature perturbation $\zeta_{k,(2)}^{\rm WKB}$ arising from Eq.~(\ref{eq:MSn}) reads
$$|\zeta_{k,(2)}^{\rm WKB}|=|v_{k,(2)}^{\rm WKB}/A|=1/\left(A\sqrt{2\bar{k}}\right)\,.$$
Then, we can express the dimensional curvature perturbation in terms of the dimensionless one through

\begin{equation}
|\zeta_k^{\rm WKB}| = \frac{A l_P}{a} \sqrt{\frac{8 \pi c_s \bar{k}}{3(1+w)k}}|\zeta_{k,(2)}^{\rm WKB}|\,.
\label{WKBzn}
\end{equation}

This relation of proportionality must be valid always, hence we can write the physical power spectrum for the dimensional curvature perturbation $\zeta_k$ in terms of the power spectrum for the dimensionless curvature perturbation $\zeta_{k,(2)}$ calculated from Eq.~(\ref{eq:MSn}):

\begin{equation}
P_\zeta=\frac{k^3}{2 \pi^2} |\zeta_k|^2 = \frac{y^2}{36 \pi c_s (1+w)}
\frac{l_P^2}{l_c^2}\bar{k}^3|\zeta_{k,(2)}|^2\,.
\label{PS}
\end{equation}

Let us evaluate semi-analytically $|\zeta_{k,(2)}|$. First of all, in the de Sitter expansion, the vacuum initial condition we have set is the well known Bunch-Davies vacuum associated with this spacetime,

\begin{equation}
v_{k,(2)} =\frac{\mathrm{e}^{-\imathnew \bar{k} \bar{\eta}}}{\sqrt{2\bar{k}}} \left(1-\frac{\imathnew}{\bar{k} \bar{\eta}}\right)\,.
\label{BD}
\end{equation}

This solution is valid whenever the de Sitter expansion is taking place, even when $\bar{k} \bar{\eta} \ll 1$, where the super-Hubble expansion \cite{Pinto-Neto:2021gcl} is also valid:

\begin{eqnarray}
\zeta_{k,(2)} &=& A_1(\bar{k})\left[1-\int\frac{\dif\bar{\eta}}{A^2}\int \dif{\bar{\eta}}_2A^2{\bar{k}}^2+\mathcal{O}({\bar{k}}^4)+\ldots\right] + \nonumber \\ &&A_2(\bar{k})\left[\int\frac{\dif\bar{\eta}}{A^2}+\mathcal{O}({\bar{k}}^2)+\ldots\right]\,.
\label{superH}
\end{eqnarray}

Comparing Eq.~(\ref{BD}) for $\bar{k} \bar{\eta} \ll 1$ with Eq.~(\ref{superH}), knowing that $\zeta_{k,(2)}=v_{k,(2)}/A$, we can evaluate the amplitude of the dominant constant mode $A_1(\bar{k})$, yielding

\begin{equation}
P_\zeta=\frac{k^3}{2 \pi^2} |\zeta_{k}|^2 = \frac{1}{16 \pi c_s}
\frac{l_P^2}{l_c^2}\left(1+\bar{k}^2 \bar{\eta}^2+\ldots\right)\,,
\label{PS2}
\end{equation}
where we substituted $w=1/3$.

Hence, we get the amplitude of the power spectrum and its spectral index, which is $0$ assuming the ideal case $w=1/3$. 

Note that in the numerical calculation of $v_{k,(2)}$, the result can only depend on $\bar{k}$ and $\bar{\eta}$, nothing more. The appearance of $y^2$ in Eq.~(\ref{PS}) was canceled because
$|\zeta_{k,(2)}|^2=|v_{k,(2)}/A|^2=|v_{k,(2)} 2 \bar{\eta}/y|^2$. Again, the presence of a very large $y$ does not pose any problem to the model, it in fact helps the calculations.

\section{Cosmological perturbations: Numerical results}
\label{secnumerical}

The analytical results obtained in Section \ref{secanalytical} can be confirmed by means of a numerical analysis, which is detailed in what follows.

In order to perform the numerical integration, the initial condition can be given as the adiabatic vacuum at very negative $\bar{\eta}$. However, in order to speed up the computation, we obtain corrections to the adiabatic vacuum and set the initial condition at not so large $|\bar{\eta}|$. This adiabatic expansion is made by firstly expanding the potential $a''/a$, equation \eqref{Vap}, for $-y\ll \bar{\eta}$, 
\be
V_{\mathrm{qdS}}\equiv \frac{1}{(1+\bar{\eta}^2)^{3/2}(\bar{\eta} + \sqrt{1+\bar{\eta}^2})} \approx \frac{2}{\bar{\eta}^{2}}-\frac{5}{2\bar{\eta}^{4}}+\mathcal{O}\left(\bar{\eta}^{-6}\right)\,,
\label{qdSpotential}
\ee
where the second approximation is valid for $|\bar{\eta}|\gg 1$. Following the procedure described in \cite{Chung:2003wn}, we obtain the initial condition
\be
  v_{k0}^{\mathrm{qdS}}=\frac{\mathrm{e}^{-\imathnew\bar{k}\bar{\eta_i}}}{\sqrt{2\bar{k}}}\left[1-\frac{\imathnew}{\bar{k}\bar{\eta_i}}+\frac{5\imathnew \bar{k}^2}{12}\frac{1}{(\bar{k}\bar{\eta_i})^3}+\mathcal{O}\left(\bar{\eta}_i^{-4}\right)\right] \,.
  \label{adiabatic_expansion_qdS}
\ee

\begin{figure}[t]
    \centering
    \includegraphics[scale=0.6]{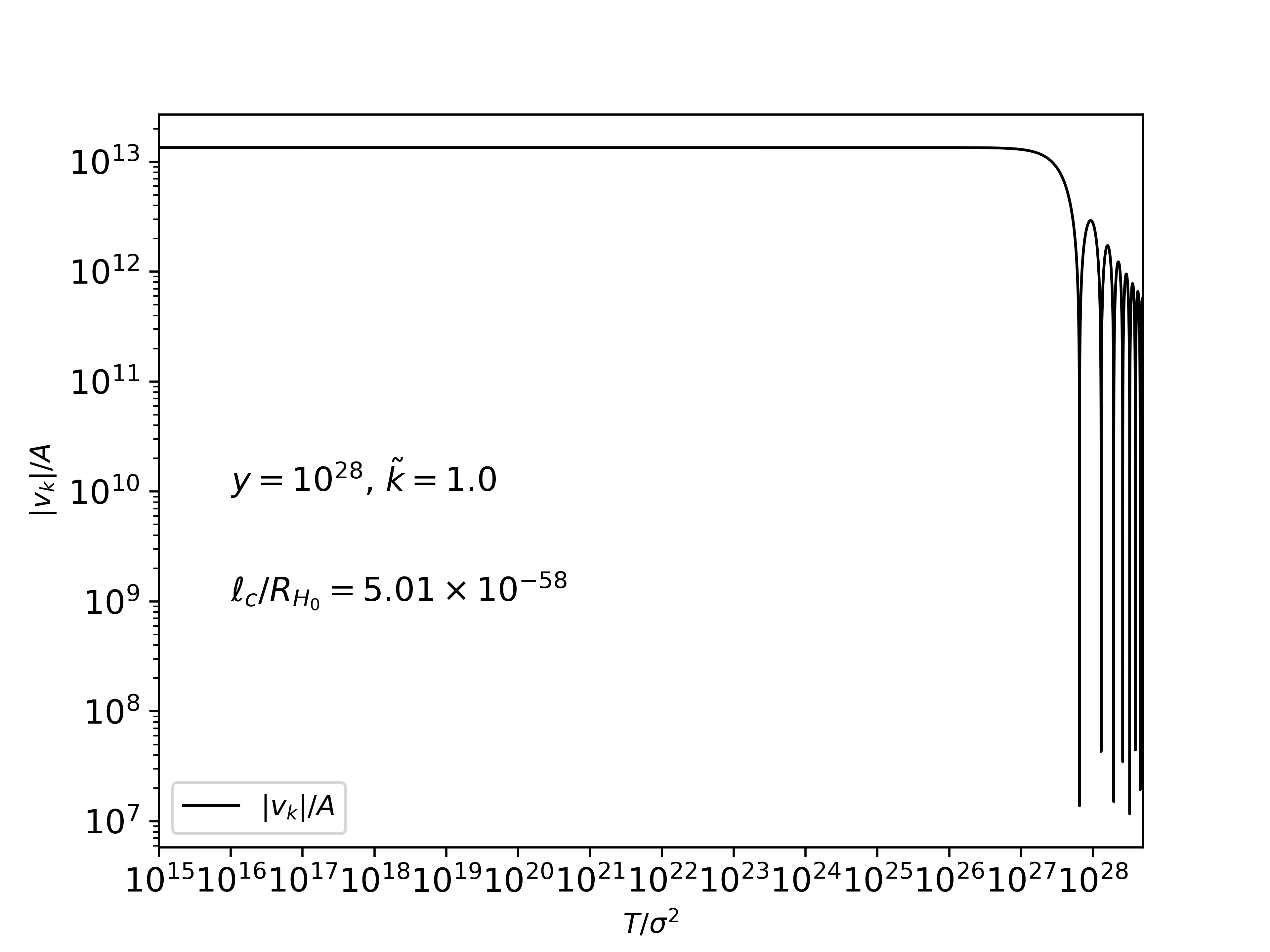}
    \caption{Modulus of the curvature perturbation mode $\zeta_k\equiv v_k/A$ for $w=1/3$, showing the horizon reentry.}
    \label{figpertnew_v/a}
\end{figure}

The first correction to the adiabatic vacuum is just the de Sitter term, see Eq.~\eqref{BD}, as expected. We then solve equation \eqref{eq:MSn} with the potential \eqref{qdSpotential} and the above initial condition at a negative time $|\bar{\eta}_i|\ll y$ set before the crossing $\bar{k}^2=V_{\mathrm{qdS}}(\bar{\eta}_c)$. 

The computational time is significantly reduced by using action-angle variables \cite{Celani:2016cwm}, which we denote by $\theta$, $I$, $\psi$ and $J$. In terms of these variables, the equation of motion  \eqref{mukhanov-sasaki1.5} reads
\begin{eqnarray}
  {\theta}'&=&\bar{k}-\frac{1}{\bar{k}}\frac{{A}''}{A}\sin^2\theta\,, \nonumber\\
 {(\ln I)}'&=&\frac{1}{\bar{k}}\frac{{A}''}{ A}\sin{(2\theta)}\,, \nonumber   \\
  {\psi}'&=&\bar{k}-\frac{1}{\bar{k}}\frac{{A}''}{ A}\sin^2\psi\,, \nonumber  \\
    {(\ln J)}'&=&\frac{1}{\bar{k}}\frac{{A}''}{A}\sin{(2\psi)}\,,
\end{eqnarray}

while the initial condition \eqref{adiabatic_expansion_qdS} becomes

\begin{eqnarray}
  \tan{\theta_i} &=& \frac{\bar{k} q_1}{{q_1}'}\,, \quad I_i = \frac{\bar{k} q_1^2}{2}+\frac{{q_1}'^2}{2\bar{k}}\,, \nonumber  \\ 
\tan{\psi_i} &=& \frac{\bar{k} q_2}{{q_2}'}\,, \quad J_i = \frac{\bar{k} q_2^2}{2}+\frac{{q_2}'^2}{2\bar{k}} \,,
\end{eqnarray}
where
\begin{equation}
\ q_1\equiv -2 \Im\left[v_{k0}^{\mathrm{qdS}}\right]\,, \quad
    q_2\equiv 2 \Re\left[v_{k0}^{\mathrm{qdS}}\right] \,.
    \label{icadiabatic}
\end{equation}
The usual Mukhanov-Sasaki variable and its derivative are then recovered as
\begin{eqnarray}
\nonumber    v_{k,(2)}=\frac{1}{\imathnew\sqrt{2\bar{k}}}\left(\sqrt{I}\sin{\theta}+\imathnew\sqrt{J}\sin{\psi}\right)\,,\\
    {v_{k,(2)}}'=\frac{1}{\imathnew}\sqrt{\frac{\bar{k}}{2}}\left(\sqrt{I}\cos{\theta}+\imathnew\sqrt{J}\cos{\psi}\right)\,.
\end{eqnarray}\\

\begin{figure}[t]
    \centering
    \includegraphics[scale=0.6]{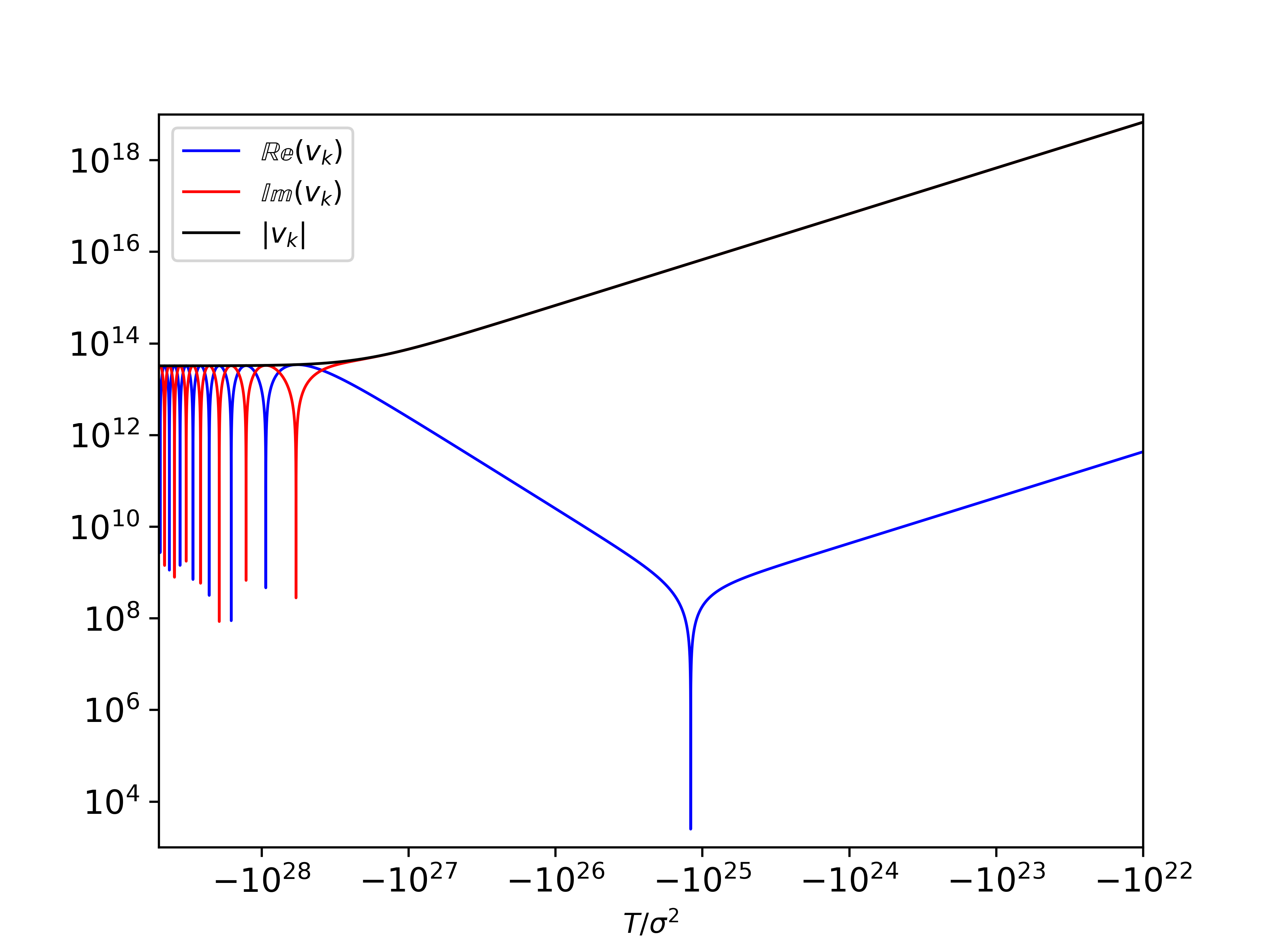}
    \caption{Absolute value of the Mukhanov-Sasaki variable $|v_{k,(2)}|$ related to the scalar perturbations, as well as its real and imaginary parts, for $y=\num{5e29}$, $l_c/l_P=\num{2.868e3}h^{-1}$ and $k_H=1.0$. }
    \label{figpertnew_vImRe}
\end{figure}

Since the qdS potential does not depend on $y$, the crossing condition leads to the usual relation $\bar{k}\bar{\eta}\approx 1$ (which corresponds precisely to $c_sk\eta\approx 1$), provided that the modes cross the potential not too close to $\bar{\eta}\approx-1$. This is the case for the frequencies \eqref{ktil2} satisfying \eqref{range22}. The initial time for the integration must then be in the range $-10^{28}<\bar{\eta}_i<-10^{15}$. Since this must happen after the bounce and within the qdS period, it follows that $y>-\bar{\eta}_i$.

\begin{figure*}[t]
\centering
\includegraphics[scale=0.16]{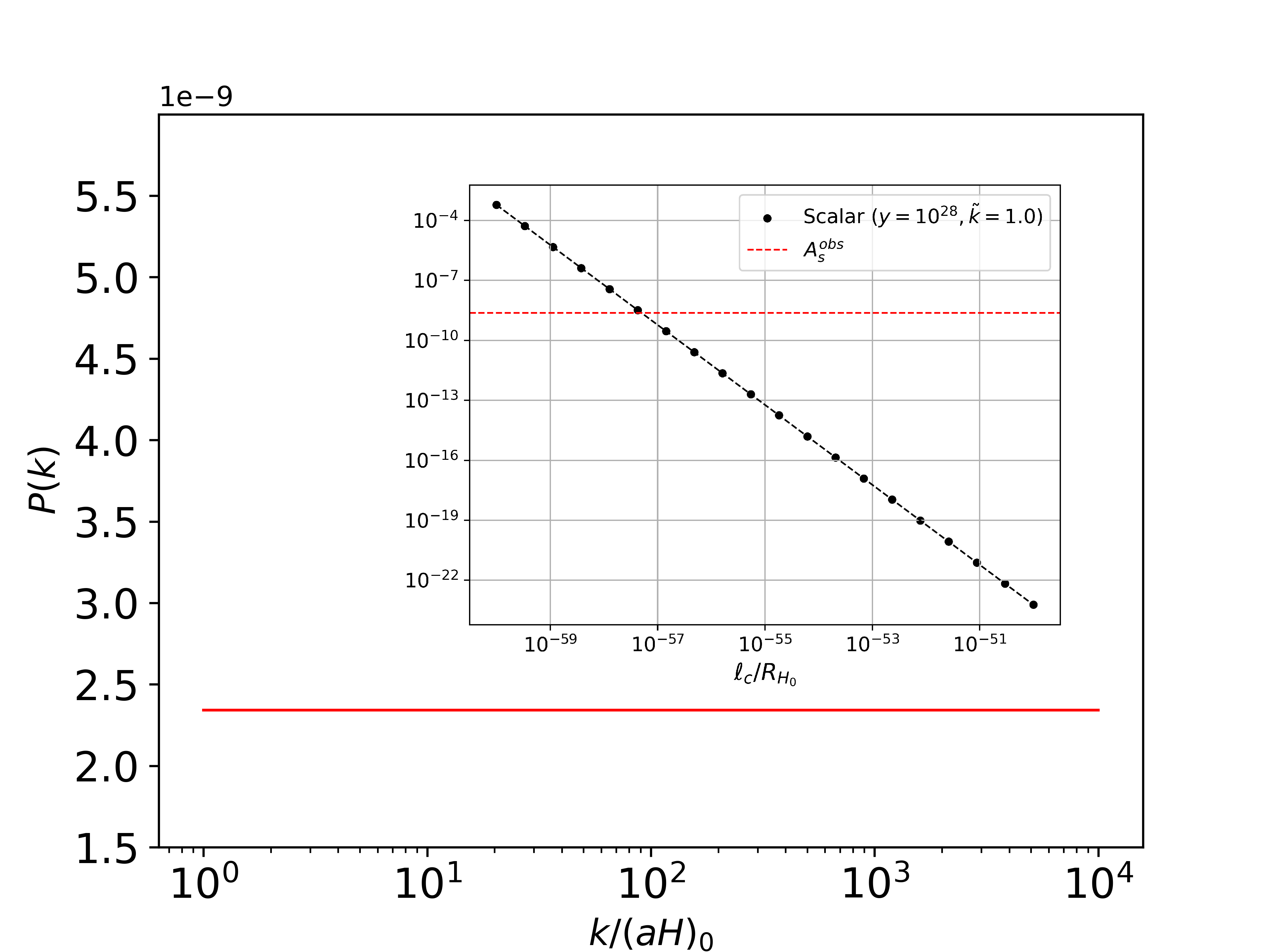}
    \caption{Scalar power spectrum obtained numerically for $w=1/3$, $l_c/l_P=\num{4.22e4}$ and initial conditions \eqref{icadiabatic} at initial time $\bar{\eta}_i=-10^{10}$. In addition, the scalar power spectrum as a function of $l_c$, with the observed scalar amplitude in red.}
\label{figpknew}
\end{figure*}

Figure \ref{figpertnew_v/a} shows the results of the numerical integration for the curvature perturbation $\zeta_k=v_{k,(2)}/a$, where one can identify the frozen regime and the oscillations after horizon reentry (for a mode with $k_H=1.0$). The initial time of integration was set at $\bar{\eta}_i=-10^{15}$, as the initial condition \eqref{adiabatic_expansion_qdS} remains appropriate even after the crossing time for this case, $\bar{\eta}_c\approx\num{-3e27}$. The real and imaginary parts of $v_{k,(2)}$ (their absolute value), as well as $|v_{k,(2)}|$, are displayed in Figure \ref{figpertnew_vImRe}. The initial adiabatic oscillations $\mathrm{e}^{-\imathnew\bar{k}\bar{\eta}}$ change, after crossing the horizon, to a growing mode proportional to the scale factor, being dominated by the imaginary part.

Now let us use the numerical results for $\zeta_{k,(2)}$ to calculate the scalar power spectrum given in Eq.~(\ref{PS}) expressed in terms of $k_H=R_{H_0}/\lambda_{\rm {physical,0}}$ for $w=c_s^2=1/3$,
\be
P_\zeta=\sqrt{\frac{2}{3}}\frac{y^2}{\pi(3\Omega_{r,0})^{3/4}}\left(\frac{l_P}{R_{H_0}}\right)^2\sqrt{\frac{R_{H_0}}{l_c}}\;k_H^3\frac{|v_{k,(2)}|^2}{A^2}\,.
\label{scalar_power_spectrum_v}
\ee

Our analytical estimates can be read in Eq.~\eqref{PS2} for $w=c_s^2=1/3$,
\be 
P_\zeta\Big{\rvert_{\bar{k}|\bar{\eta}|< 1}}\approx \frac{\sqrt{3}}{16\pi}\left(\frac{l_P}{l_c}\right)^2\times\left[1+{\mathcal{O}}\left(\bar{k}\bar{\eta}\right)^2+\ldots\right]\,,
\label{PS3}
\ee
which of course corresponds to \eqref{scalar_power_spectrum_v} evaluated with the dominant term in \eqref{adiabatic_expansion_qdS} at horizon crossing.
We computed the scalar power spectrum numerically from the perturbation amplitude on super-horizon scales by means of expression \eqref{scalar_power_spectrum_v}, which is shown in figure \ref{figpknew}. Note that the dependence on $l_c^{-2}$ given in Eq.~\eqref{PS3} is confirmed. We also went beyond the observational range, and deviations from scale invariance only occur close to the cutoff scale \eqref{cutoff_scale}. The scalar amplitude compatible with the CMB observations $A_s=\num{2.3424e-9}$, where $P_\zeta=A_s(k/k_*)^{n_s-1}$, leads to the numerical value $l_c/R_{H_0}=\num{5.01e-58}$, which in turn translates to
\be
\frac{l_c}{l_P}=\num{2.868e3}h^{-1}\,.
\label{lc_numerical_value}
\ee
As discussed in Section \ref{secBackEv}, this is a very reasonable value (see Eq.~\eqref{lc_bounds}), taking the usual CMB/Supernovae values $0.67<h<0.74$.

From this value of the minimum curvature scale we obtain through the energy density provided by \eqref{effective_Lambda_qdS} the characteristic energy scale in the qdS phase
\be
H^*=\sqrt{\frac{\Lambda_{\mathrm{qdS}}}{3}}\approx 10^{14}\,\mathrm{GeV}\,,
\label{energy scale inflation}
\ee
which is similar to that of usual inflationary models. 

It was also noticed that the power sepctrum does not depend on $y$ for $y\gg 1$. However, since $y>C$ and from Eq.~\eqref{range22} , we obtain that

\be
 y\gtrsim 10^{27}\,,
 \label{y_bound_Pzetafinal}
\ee

which is in remarkable agreement with the inflationary e-folds lower bound \eqref{e-folds bounds} (note that $y$ can be even larger, the only effect being pushing the start of the qdS phase farther into the past). The observational constraints are thus contained in the allowed parameter space of this model.

We observe that, as expected for the case $w=1/3$, the power spectrum is very nearly scale invariant, with a deviation from $n_s=1$ in the fifth decimal place (this makes the normalization by the pivot scale $k_*$ of little effect to the amplitude of $A_s$).

The cutoff scale \eqref{potential_cutoff} may now be evaluated quantitatively. From $\bar{k}_{\mathrm{max}}^2=V(\bar{\eta}_{\mathrm{max}})$ we find 
\be
{k_H}_\mathrm{max}\approx\num{1.6e27}\; \to \; k_\mathrm{max}\approx\num{5.3e20}h\,\mathrm{Mpc}^{-1}\,,
\label{cutoff_scale}
\ee
being well beyond the modern observational limits. Any perturbation mode above this value will never cross the horizon, remaining ``sub-Hubble'' throughout the entire evolution of the universe. For $h\simeq 0.67$ they correspond to $\lambda_{\mathrm{physical},0}\approx 0.54\;\mathrm{m}$, which at the time of perturbation generation would be below the Planck scale.

In summary, the numerical results are consistent with previous analytical estimates and the amplitude of almost scale invariant scalar perturbations are in agreement with current observational constraints, indeed mimicking inflation. The fact that a qdS phase can be achieved due to quantum effects, both at the background and perturbation levels, is remarkable. 

\section{Conclusion}
\label{conc}

In this paper we presented a very simple nonsingular cosmological model which contains a classical slowly contracting phase, a bounce, a quasi-de Sitter inflationary phase, and finally reaches the usual classical radiation-dominated expanding phase before nucleosynthesis. The unique matter component is a radiation fluid. The bounce and the inflationary phase result from quantum effects arising from a Gaussian wave function solution of the Wheeler-De Witt equation, which reduces to a Schr\"odinger  equation in this case, travelling with high momentum in configuration space. During the quantum phase, there is a huge creation of photons, so that the contracting phase corresponds to an almost empty universe.

All this rich phenomenology is described by the astonishingly simple analytical scale factor given in Eq.~\eqref{eq2:aw=1/3} evolving in conformal time:
\begin{equation}
    a(\bar{\eta}) = a_b \left( y\bar{\eta}  + \sqrt{1 + y^2} \sqrt{1 + \bar{\eta}^2} \right) \,,
    \label{eq2:aw=1/3-2}
\end{equation}
Independently of its origin, Eq.~\eqref{eq2:aw=1/3-2} is a new, so far unknown, scale factor evolution which is amazingly interesting in itself, and it is really worth looking for other theoretical contexts where it can be obtained. For instance, the classical contracting phase happens with a very tiny number of photons in a given cell with physical volume $V_{\rm phys} = a^3 V_{\rm com}$, where $V_{\rm com}$ is the comoving volume of the cell,

\begin{equation}
\frac{N_-(a)}{N_+(a)} \approx \frac{1}{16 y^4} < 10^{-109} \,,
\label{ratioNumber}
\end{equation}
where $N_{\pm}(a)$ is the total number of photons in the same cell with a given $a$ at the classical expanding (contracting) phase, respectively. In order to obtain Eq.~\eqref{ratioNumber}, we used Eqs.~ \eqref{ratio2} and \eqref{y_bound_Pzetafinal}. Taking the cell to have the volume of our universe today, which contains around $10^{90}$ photons, the same volume in the classical contracting phase would have $10^{-19}$ photons or less, so it was practically empty\footnote{Note that the scale factor corresponding to today's Hubble radius yields a much bigger Hubble radius in the classical contracting phase, see Eq.~\eqref{ratio3}, compatible with the fact that spacetime in this era is close to Minkowski spacetime.}. Hence, one may think that the scale factor \eqref{eq2:aw=1/3-2} might have been originated from some quantum gravity instability of a primordial Minkowski spacetime, leading to a tiny number of massless particles, which is substantially increased at the bounce and inflationary quantum phases afterwards.

Note that the generalization of Eq.~\eqref{eq2:aw=1/3-2} to any $w=p/\rho=$const. is also very simple, see Eq.~\eqref{scale_factor}, where $\dif\bar{T}=a^{1-3w} \dif\bar{\eta}$, allowing many different scenarios and possibilities. For instance, in the case of the matter bounce scenario, $w\approx 0$, there is a phantom-like expanding phase after the bounce without any phantom, the consequences of which might be interesting to be investigated. 

Taking the background as given by Eq.~\eqref{eq2:aw=1/3-2} for $y\gg 1$, we found out that the scalar cosmological perturbations are almost scale invariant and with the right amplitude for the scales observed in the Planck satellite \cite{Planck:2018jri}. The value of $y$ for which these observed scales acquire their observed properties should satisfy the inequality $y\gtrsim 10^{27}$, which coincides with a number of e-folds $\mathcal{N}$ during inflation given by  $\mathcal{N} > 60$, see \eqref{e-folds bounds}. Furthermoe, the observed value of the amplitude of scalar perturbations by Planck imposes that the minimum curvature scale of the background model $l_c$ should be 
\be
\frac{l_c}{l_P}=\num{2.868e3}h^{-1}\,,
\ee
where $l_P$ is the Planck length, which is consistent with the quantum approach we are using: it is not of the order of the Planck length, where a yet unknown full theory of quantum gravity should be used inescapably, but it still corresponds to energy scales far bigger than that of nucleosynthesis, where quantum cosmological effects may begin to be important. It was also obtained that the energy scale of the quantum inflationary phase is about $E_{\rm qi}\approx 10^{14}\,\mathrm{GeV}$,
see Eq.~\eqref{energy scale inflation}.

Concluding, this astonishing simple non-singular cosmological model yields many observed features of the standard cosmological model, and it naturally contains three key ingredients of the primordial universe which have been investigated so far: a very slow contraction (although with a different origin than that of the usual ekpyrotic scenarios), a bounce, and a inflationary phase, reaching the standard classical radiation-dominated phase before nucleosynthesis. All these different phases are continuously connected within the simple expression given in Eq.~\eqref{eq2:aw=1/3-2}. 

Of course, this very simple model is an important first step, but it is not the final word, and it must be supplemented by new ingredients. For instance, and perhaps most importantly, it does not lead to a red-tilted spectral index, unless we take $w=1/3+\epsilon$ with $0<\epsilon\ll 1$ in order to give $n_s \approx 0.9649$ \cite{Planck:2018jri} (see the analytic result \eqref{Pzeta_w_dep}), nor to primordial gravitational waves with the right amplitude. In order to understand this last point, take the tensor metric perturbations $w_{ij}=\zeta^h e_{ij}$, where $e_{ij}$ is the transverse-traceless polarization 3-tensor \cite{Peter_etal_Tensor}. It can be shown that the perturbation mode $v_{k}^h=a\zeta^h_k$ satisfies equation \eqref{mukhanov-sasaki1} with $c_s=1$. After calculations similar to those employed in the scalar case, we find that the tensor power spectrum $P_h$ will also be scale invariant for modes that cross the horizon during the qdS phase, but with the tensor-to-scalar ratio given by

\be
r\equiv \frac{A_t}{A_s}\simeq\frac{32}{\sqrt{3}}\approx 18.48\,.
\label{r_analytical}
\ee
where $P_h=A_t(k/k_*)^{n_t}$, with $k_*$ a selected pivot scale. 

Complete numerical computations are in full agreement with the analytical predictions given in \eqref{r_analytical}, to a percentage error of around $10^{-6}$. This clearly violates the observational constraint $r<0.063$ \cite{Planck:2018jri}. Note that this specific issue is not present in a fluid matter bounce model, as shown in Ref.~\cite{PhysRevD.78.063506} for a symmetric bounce. 

One way out usually implemented to solve these types of problems is to evoke the presence of a curvaton field (maybe associated to dark matter, which is here absent) \cite{Gordon:2000hv, Gong:2016yyb,Lyth:2003ip,Cai:2011zx}, which does not affect the background evolution, but whose presence can increase the amplitude of scalar perturbations with respect to tensor perturbations. Also, one can induce a red-tilt in the spectrum index of scalar perturbations by considering an effective global equation of state parameter $w=1/3+\epsilon$, as mentioned above, where the small deviation is due to the presence of the curvaton field. These are issues to be investigated in a forthcoming paper.

\begin{acknowledgments}

PAPM acknowledges financial support from CAPES (Brazil) under grant No. 88887.666393/2022-00. PCMD is supported by the grant No. UMO-2018/30/Q/ST9/00795 from the National Science Centre, Poland. RFP and NPN acknowledge
the support of CNPq of Brazil under grants 140696/2022-9 and PQ-IB
310121/2021-3, respectively.
\end{acknowledgments}

\bibliography{main} 

\end{document}